\documentclass[aps]{revtex4}

\begin{document}

\title{Asymptotic Expansions of Feynman Amplitudes in a Generic Covariant Gauge}
\author{C.A. Linhares$^{a}$,  A.P.C. Malbouisson$^{b}$ and I. Roditi$^{b}$}

\address{$^{a}$Instituto de F\'{\i}sica, Universidade do Estado do Rio de Janeiro,\\
Rua S\~{a}o Francisco Xavier, 524, 20559-900 Rio de Janeiro, RJ, Brazil}

\address{$^{b}$Centro Brasileiro de Pesquisas F\'{\i}sicas,\\
Rua Dr.~Xavier Sigaud, 150, \\
22290-180 Rio de Janeiro, RJ, Brazil}

\begin{abstract}
We show in this paper how to construct Symanzik polynomials and the
Schwinger parametric representation of Feynman amplitudes for gauge theories
in a generic covariant gauge. The complete Mellin representation of such
amplitudes is then established in terms of invariants (squared sums of
external momenta and squared masses). From the scaling of the invariants by
a parameter we extend for the present situation a theorem on asymptotic
expansions, previously proven for the case of scalar field theories, valid
for both ultraviolet or infrared behaviors of Feynman amplitudes.
\end{abstract}

\maketitle

\section{Introduction}

Asymptotic expansions for amplitudes related to Feynman diagrams continue to
be a topic of interest. This is specially true in what regards the infrared
and ultraviolet behaviors of gauge theories, studies of critical phenomena
and attempts to understand confinement in the field-theoretical framework of
quantum chromodynamics. With respect to gauge theories, whose interaction
mediator field has a propagator of the form (in Euclidean space) 
\begin{equation}
\frac 1{k^2+\mu ^2}\left( \delta _{\mu \nu }-(1-\xi )\frac{k_\mu k_\nu }{%
k^2+\mu ^2}\right) ,  \label{propagator}
\end{equation}
where $\mu $ is an infrared mass regulator and $\xi $ is the gauge-fixing
parameter, amplitude computations are usually performed by choosing, for
simplicity, the ($\xi =1$) Feynman gauge. This has been done, for example,
in a recent paper directly connected to the subject of our interest \cite
{derafael}, where its authors were able to produce asymptotic expansions for
some 2- and 3-loop vertex corrections in QED, using a method based on the
Mellin--Barnes transform of Feynman amplitudes. There are, however, some
circumstances in which it is desirable to work with other choices of $\xi $,
normally involving non-gauge-invariant quantities. This is the case, for
instance, of recent discussions \cite{gribov, silvio} on the form factors of
gluon and ghost propagators in infrared QCD, whose computations cannot be
formulated in the Feynman gauge. Also, in the context of noncommutative
supersymmetric field theories, some calculations of radiative corrections
require the choice of a non-Feynman gauge \cite{ferrari}. The intrinsic
complexity of working in a general gauge, and its relevance for physical
situations as those just mentioned, have led us to consider establishing
asymptotic expressions in a general \emph{covariant }regime. We shall do
this by using the method of the complete Mellin representation of Feynman
amplitudes. It shares with the Mellin transform method of Ref. \cite
{derafael} the advantage of avoiding the explicit calculation of amplitudes
before deriving the asymptotic expressions.

Thus, in the present paper, we wish to extend the method when one considers
the amplitudes in a generic covariant gauge. We here intend to take up the
long tradition that started in the 1970's with the papers \cite{bergere3,
bergere1, bergere2} within the Bogoliubov--Parasiuk--Hepp--Zimmermann
formalism. Their approach is based on the Schwinger representation of
amplitudes. This representation has the advantage of being expressed in
terms of so-called Symanzik polynomials in the Schwinger parameters, which
can be deduced in a straightforward way from the topological characteristics
of the diagrams one is interested in. From the topological point of view,
the diagrams are considered as a set of interconnected lines, ignoring
distinctions between those related to gauge fields and those to matter
fields. These `topological formulas' are thus determined from inspection of
the 1-trees and 2-trees constructed from the diagram by omitting the
adequate number of lines \cite{nakanishi, itzykson}. The Mellin transform
technique was then used to prove theorems leading to the asymptotic
expansion of convergent Feynman diagrams.\ External momenta are scaled by
some parameter $\lambda $ and through the Mellin transform with respect to
the scaling parameter, as one takes the limit of $\lambda $ going to
infinity, the existence of an asymptotic series in powers of $\lambda $ and
powers of logarithms of $\lambda $ is proved, and the series coefficients
computed. Divergent diagrams were treated in \cite{bergerelam}.

In the subsequent years, the subject had further advancements, with the
inclusion of the internal squared masses as possible invariants which may be
scaled, in order to study the infrared behavior of the amplitudes as well 
\cite{malbouisson7}. There the concept of `FINE' polynomials was introduced,
that is, those having the property of being factorizable in each Hepp sector 
\cite{hepp} of the variables. It was then argued that they may be
`desingularized', which means that the integrand has a meromorphic
structure, so that the residues of its various poles may be obtained.
However, this is not the general situation, and in many diagrams the FINE
property does not occur, which is required to hold by the Mellin transform
method. A first solution to this problem was presented in Ref.~\cite
{malbouisson7} by introducing the so-called `multiple Mellin'
representation, which consists essentially in splitting the Symanzik
polynomials in a certain number of pieces, each one of which having the FINE
property. Then, after scaling by the parameter $\lambda $, an asymptotic
expansion can be obtained as a sum over all Hepp sectors. This is always
possible to do, if one adopts, as was done in Refs.~\cite{malbouisson,
rivasseau, malbouisson5}, the extreme point of view which consists in
splitting the Symanzik polynomials in all its monomials. Moreover, this
apparent complication is compensated by the fact that one can dispense with
the use of Hepp sectors altogether. This came to be known as the `complete
Mellin' (CM) representation.

The CM representation provides a general proof of the existence of an
asymptotic expansion in powers and powers of logarithms of the scaling
parameter and no change of variables is needed: not only the Schwinger
variables provide a desingularization, but the integrations over such
variables can be explicitly performed, and we are left with the pure
geometrical study of convex polyhedra in the Mellin variables. It has been
proven \cite{malbouisson} that the same results of \cite{malbouisson7} are
obtained in a simpler way, and asymptotic expansions are computed in a more
compact form, without any division of the integral into the Hepp sectors.
Another feature of the CM representation is that it allows a unified
treatment of the asymptotic behavior of both ultraviolet convergent and
divergent amplitudes. Indeed, as is shown in Refs. \cite{malbouisson,
rivasseau}, the renormalization procedure does not alter the algebraic
structure of integrands in the CM representation. It only changes the set of
relevant integration domains in the Mellin variable space.

Other applications of the method included a study of the infrared behavior
of amplitudes relevant to critical phenomena \cite{malbouisson6} and of
dimensional regularization \cite{malbouisson5}.

In Sec. II of the present paper, we show how to write down a Schwinger
representation for a Feynman amplitude, when the gauge-fixing parameter $\xi 
$ is left unchosen, through the use of a `topological formula'. In this
case, then, as seen in eq. (\ref{propagator}), every gauge propagator in a
diagram contributes with a second term with momentum dependence $(k^2+\mu
^2)^{-2}$, in addition to the $(k^2+\mu ^2)^{-1}$ dependence of the first
term, the latter one being the same that occurs in scalar theories and gauge
theories in the Feynman gauge. This fact entails the introduction of two
Schwinger parameters to the same gauge-field internal line of the diagram.
Still, one should consider the momentum dependence occuring in the
numerator, coming not only from the gauge-field propagator, but also from
those of fermionic particles, and, in principle, from momentum-dependent
vertices and counterterms. Numerators, however, modify the topological
formulas in a systematic way \cite{bergere3, bergere2, rivasseau}.

Sec. III reviews the complete Mellin transform method, and we demonstrate in
Sec. IV the existence of the asymptotic series for gauge theories in our
generalized case. In Sec. V we present our conclusions.

The Euclidean metric will be used throughout the paper.

\section{Parametric representation and topological formulas}

We start by considering an arbitrary connected Feynman diagram $G$ in a
purely scalar theory and let us denote by $I_G$ its contribution to some $n$%
-point Green's function in momentum space. It is a function of the
relativistic invariants built from the external momenta and squared masses.
Topologically, the Feynman diagram $G$ is a set of $I$ internal lines, $V$
vertices and $L$ loops and these quantities are well-known to satisfy the
relation $L=I-V+1$. A $q$-tree of the diagram $G$ is a subdiagram of $G$
having $q$ connected components, without loops, and linking all the vertices
of $G$. Of particular interest for us are the cases of $q=1$ (1-trees) and $%
q=2$\ (2-trees).

Let us define $P_v$ as the sum of all external momenta that end at the
vertex $v$ and let us assume that they are such that $\sum_{v=1}^VP_v=0$.
Apart from global momentum conservation and overall constant factors coming
from the vertices (for instance, such as $\left( -i\lambda \right) ^V$ in $%
\lambda \phi ^4/4!$ theory) or from a possible symmetry factor, the
arbitrary amplitude reads 
\begin{equation}
I_G(P_v;m_l^2)=\int \prod_{l=1}^I\frac{d^dk_l}{(2\pi )^d}\frac
1{k_l^2+m_l^2}\prod_{v=1}^V(2\pi )^d\delta ^d\left( P_v-\sum_{l=1}^I\epsilon
_{vl}k_l\right) .  \label{amplitude1}
\end{equation}
In this formula, we integrate the Euclidean propagators (with possibly
different masses $m_l$) over all momenta $k_l$ associated to each internal
line $l$. The last factor ensures momentum conservation at each vertex. In
it we have introduced the so-called incidence matrix, which is defined as 
\begin{equation}
\epsilon _{vl}=\left\{ 
\begin{array}{ll}
+1 & \text{if the vertex }v\text{ is the starting point of line }l \\ 
-1 & \text{if the vertex }v\text{ is the ending point of line }l \\ 
0 & \text{otherwise.}
\end{array}
\right.
\end{equation}

The Schwinger parametric integral $\alpha $-representation is introduced by
expressing each propagator in the form 
\begin{equation}
\frac 1{k^2+m^2}=\int_0^\infty d\alpha \,e^{-\alpha (k^2+m^2)}
\label{schwinger}
\end{equation}
and the delta function is expressed as an integral as well, 
\begin{equation}
(2\pi )^d\delta ^d\left( P_v-\sum_{l=1}^I\epsilon _{vl}k_l\right) =\int
d^dy\,e^{-iy\cdot \left( P_v-\sum_l\epsilon _{vl}k_l\right) }.
\end{equation}
Upon integration over each $k_l$ and then over the $y$ variables, the
amplitude is finally expressed as \cite{nakanishi, itzykson} 
\begin{equation}
I_G(P_v;m_l^2)=\int_0^\infty \frac{\prod_{l=1}^Id\alpha _l}{(4\pi
)^{dL/2}U^{d/2}(\alpha )}e^{-\sum_l\alpha _lm_l^2}e^{-N(s_K;\alpha
)/U(\alpha )},  \label{amplitude2}
\end{equation}
where $U$ and $N$ are homogeneous polynomials in the $\alpha _l$ parameters,
of order $L$ and $L+1$, respectively, constructed with topological relations
defined by the 1- and 2-trees of the diagram $G$: 
\begin{equation}
U(\alpha )=\sum_T\prod_{l\notin T}\alpha _l,\qquad N(\alpha
)=\sum_Ks_K\left( \prod_{l\notin K}\alpha _l\right) .  \label{polynomials}
\end{equation}

The symbols $\sum_T$ and $\sum_K$ denote, respectively, summation over the
1-trees $T$ and 2-trees $K$ of $G$. $s_K$ is the cut-invariant of one of the
two connected pieces of the 2-tree $K$, that is, the square of the total
external momentum $P_v$ entering one of pieces of $K$ (any one of them
equivalently, by momentum conservation). $U$ and $N$ are known in the
literature as the Symanzik polynomials.

Let us consider now a Yukawa-type scalar--fermion theory. A generic
amplitude contains a certain number of fermion propagators which introduce a
momentum-dependent polynomial $\mathcal{P}(k_l)=\prod_l(\not{k}_l-m_f)$ as a
numerator in its expression, $m_f$ being the fermion mass. Differently from
eq. ~(\ref{amplitude1}), the amplitude now is expressed as 
\begin{equation}
I_G(P_v;m_l^2)=\int \left( \prod_{l=1}^I\frac{d^dk_l}{(2\pi )^d}\right) 
\frac{\mathcal{P}(k_l)}{k_l^2+m_l^2}\prod_{v=1}^V(2\pi )^d\delta ^d\left(
P_v-\sum_{l=1}^I\epsilon _{vl}k_l\right) .
\end{equation}
The internal line index $l$ runs indistinctively over scalar or fermionic
lines and $m_l$ is to be taken either as the scalar field mass or the
fermion mass, accordingly. If we take the integral representation \cite
{bergere3, bergere2, rivasseau} 
\begin{equation}
\frac{\mathcal{P}(k_l)}{k_l^2+m_l^2}=\left. \int_0^\infty d\alpha _l\,%
\mathcal{P}\left( -\frac 1{\alpha _l}\frac \partial {\partial \zeta
_l}\right) e^{-\alpha _l\left( k_l^2+m_l^2+k_l\cdot \zeta _l\right) }\right|
_{\zeta _l=0},  \label{derivative}
\end{equation}
the integrations in the amplitude may be performed in a similar way as in
the purely scalar case, with the result that 
\begin{equation}
I_G(P_v;m_l^2)=\left. \int_0^\infty \frac{\prod_{l=1}^Id\alpha _l}{(4\pi
)^{dL/2}U^{d/2}(\alpha )}\mathcal{P}\left( -\frac 1{\alpha _l}\frac \partial
{\partial \zeta _l}\right) e^{-\sum_l\alpha _l\bar{m}_l^2}e^{-N(\bar{s}%
_K;\alpha )/U(\alpha )}\right| _{\zeta _l=0},  \label{Yukawa}
\end{equation}
where 
\begin{equation}
\bar{m}_l^2=m_l^2-\zeta _l^2/4  \label{mbar}
\end{equation}
and $\bar{s}_K$ is the same cut-invariant of the 2-tree $K$ that occurs in (%
\ref{polynomials}), for which the external momenta $P_v$ is replaced by 
\begin{equation}
\bar{P}_v=P_v+\sum_l\epsilon _{vl}\zeta _l/2.  \label{pbar}
\end{equation}

Let us now pass to our main subject, fermionic gauge theories. For a given
generic diagram $G$, let us call $F$ the set of its (fermionic)\ matter
field lines and $\Gamma $ the set of its gauge-field propagators. The set of
all lines in $G$ is denoted by $E=F\cup \Gamma $. The number of lines in
each of these sets are denoted by $N_F$ and $N_\Gamma $, respectively, and
therefore the total number of propagators is $N_E=N_F+N_\Gamma $. In an
arbitrary gauge, using the gauge-field propagator (\ref{propagator}), the
amplitude $I_G$, related to the diagram $G$, is 
\begin{equation}
I_G(\xi )=C_G\int \left( \prod_{l=1}^L\frac{d^dk_l}{(2\pi )^d}\right)
\prod_{i=1}^{N_F}\mathcal{P}(q_i)\frac 1{q_i^2+m_i^2}\prod_{j=1}^{N_\Gamma
}\left( \frac 1{p_j^2+\mu ^2}-(1-\xi )\frac{p_jp_j}{\left( p_j^2+\mu
^2\right) ^2}\right) ,  \label{IG1+IG2}
\end{equation}
which, in the simpler case of the Feynman gauge, reduces to 
\begin{equation}
I_G(\xi =1)=C_G\int \left( \prod_{l=1}^L\frac{d^dk_l}{(2\pi )^d}\right)
\prod_{i=1}^{N_F}\mathcal{P}(q_i)\frac 1{q_i^2+m_i^2}\prod_{j=1}^{N_\Gamma
}\frac 1{p_j^2+\mu ^2}.  \label{intG1}
\end{equation}
To deal with the general expression, we recast it in the form 
\begin{equation}
I_G(\xi )=C_G\int \left( \prod_{l=1}^L\frac{d^dk_l}{(2\pi )^d}\right)
\prod_{i=1}^{N_F}\mathcal{P}(q_i)\frac 1{q_i^2+m_i^2}\prod_{j=1}^{N_\Gamma }%
\mathcal{Q}(\xi ,p_j)\frac 1{\left( p_j^2+\mu ^2\right) ^2},  \label{intG2}
\end{equation}
where $\mathcal{Q}$ is defined as $\prod_{j=1}^{N_\Gamma }\left( p_j^2+\mu
^2-(1-\xi )p_jp_j\right) $, thus carrying the dependence on the gauge
choice. We see that $I_G(\xi =1)$ has a similar structure with respect to
that of the Yukawa case above. Here, we collect in $C_G$ all global factors;
the parameters $m_i$ are the (possibly different) matter-field masses and $%
\mu $ is the gauge-field infrared regulator. As before, let $P$ denote the
set of external momenta; thus, to shorten the expression, we have considered
the momenta $q_i(P,\{k_l\})$ and $p_j(P,\{k_l\})$ as linear functions of the
set of $k_l$ obtained by momentum conservation at each vertex. $\mathcal{P}$
and $\mathcal{Q}$ are polynomials in the momenta, coming from the numerators
of the propagators. They also contain factors such as the metric tensor and
Dirac matrices, appropriate for a given diagram $G$ (spacetime indices are
omitted).

For both of the integrals above, a Schwinger parametric representation can
be written down. In order to do this, we particularize eq. ~(\ref{derivative}%
) for each of the numerators in the integrals (\ref{intG1}) and (\ref{intG2}%
). Integral (\ref{intG1}) has an analogous representation like in Yukawa
theory, eq. ~(\ref{amplitude2}), in which a vector $\zeta _i$ is attached to
each fermionic line $i$ such that the momenta in the $\mathcal{P}$
polynomial are substituted by a corresponding derivative with respect to $%
\zeta _i$ at $\zeta _i$ equal to zero. We can then write the Feynman-gauge
amplitude as 
\begin{equation}
I_G(\xi =1)=C_G\left. \int_0^\infty \frac{\prod_{l=1}^{N_E}d\alpha _l}{(4\pi
)^{dL/2}U^{d/2}}\prod_{i\in F}\mathcal{P}\left( -\frac 1{\alpha _i}\frac
\partial {\partial \zeta _i}\right) e^{-\sum_{i=1}^{N_F}\bar{m}_i^2\alpha
_i-\mu ^2\sum_{j=1}^{N_\Gamma }\alpha _j}e^{-N(\bar{s}_K)/U}\right| _{\zeta
_l=0},  \label{IG1novo}
\end{equation}
with the Symanzik polynomials given by the same definitions as above, eqs.~ (%
\ref{polynomials}), modified by the presence of the auxiliary variables $%
\zeta _l$, just as in the scalar--fermion case, eqs. ~(\ref{mbar}) and (\ref
{pbar}).

For the general case, we remind that the starting point for the parametric
representation is eq.~ (\ref{schwinger}). Therefore, in eq. ~(\ref{intG2}),
as we have products of $(p_j+\mu ^2)^2$ in the denominator, we should
associate two Schwinger parameters, say, $\beta _1$ and $\beta _2$, to the
same line, that is, 
\begin{equation}
\frac 1{(p^2+\mu ^2)^2}=\left( \int_0^\infty d\beta _1\,e^{-\beta _1(p^2+\mu
^2)}\right) \left( \int_0^\infty d\beta _2\,e^{-\beta _2(p^2+\mu ^2)}\right)
.
\end{equation}
This means that the set of all lines of the diagram, related to the momenta $%
q_i$ and $p_j$, is such that to each matter-field line $i$ is attributed a
parameter $\alpha _i$, whereas for each gauge-field line $r$ is associated a
couple of parameters $\beta _{r_1}$, $\beta _{r_2}$. Thus the parametric
representation for $I_G(\xi )$ has the form 
\begin{eqnarray}
I_G(\xi ) &=&C_G\int_0^\infty \frac{\prod_{i=1}^{N_F}d\alpha
_i\prod_{r_1,r_2=1}^{N_\Gamma }d\beta _{r_1}d\beta _{r_2}}{(4\pi
)^{dL/2}U^{d/2}}  \nonumber \\
&&\times \prod_{i\in F}\mathcal{P}\left( -\frac 1{\alpha _i}\frac \partial
{\partial \zeta _i}\right) \prod_{r_1\in \Gamma }\mathcal{Q}_1\left( -\frac
1{\beta _{r_1}}\frac \partial {\partial \zeta _{r_1}}\right) \prod_{r_2\in
\Gamma }\mathcal{Q}_2\left( -\frac 1{\beta _{r_2}}\frac \partial {\partial
\zeta _{r_2}}\right)  \nonumber  \label{IG2novo} \\
&&\times \left. e^{-\sum_{i=1}^{N_F}\bar{m}_i^2\alpha
_i-\sum_{r_1=1}^{N_\Gamma }\bar{\mu}_{r_1}^2\beta
_{r_1}-\sum_{r_2=1}^{N_\Gamma }\bar{\mu}_{r_2}^2\beta _{r_2}}e^{-N(\bar{s}%
_K)/U}\right| _{\zeta =0}.  \label{IG2novo}
\end{eqnarray}
where now 
\begin{equation}
U=\sum_T\prod_{i\notin T}\alpha _i\prod_{r_1\notin T}\beta
_{r_1}\prod_{r_2\notin T}\beta _{r_2},\qquad N(\bar{s}_K)=\sum_K\bar{s}%
_K\prod_{i\notin K}\alpha _i\prod_{r_2\notin T}\beta _{r_1}\prod_{r_2\notin
T}\beta _{r_2}.
\end{equation}

\section{The Complete Mellin representation for scalar theories}

Let us first discuss the simpler case of a Feynman amplitude in a scalar
theory. The Feynman amplitude (\ref{amplitude2}) has a complete Mellin
representation \cite{malbouisson}, obtained in the following way. Let us
rewrite the Symanzik polynomials as 
\begin{equation}
U(\alpha )=\sum_j\prod_{l=1}^I\alpha _l^{u_{lj}}\equiv \sum_jU_j,\;\;\qquad
N(\alpha )=\sum_Ks_K\left( \prod_{l=1}^I\alpha _l^{n_{lK}}\right) \equiv
\sum_KN_K,  \label{sym}
\end{equation}
where 
\begin{equation}
u_{lj}=\left\{ 
\begin{array}{ll}
0 & \text{if the line }l\text{ belongs to the 1-tree }j \\ 
1 & \text{otherwise}
\end{array}
\right.
\end{equation}
and 
\begin{equation}
n_{lK}=\left\{ 
\begin{array}{ll}
0 & \text{if the line }l\text{ belongs to the 2-tree }K \\ 
1 & \text{otherwise.}
\end{array}
\right.
\end{equation}
Moreover, 
\begin{equation}
\sum_lu_{lj}=L\,;\,\;\;\;\;\sum_ln_{lK}=L+1,
\end{equation}
for all $j$ and $K$, respectively.

The idea of the complete Mellin representation has its roots in the
so-called `multiple Mellin' representation introduced in \cite{malbouisson7}%
. In this case, the polynomial $N$ in eq.~ (\ref{amplitude2}) is split into
pieces, $N=\sum_aN_a$, in such a way that each piece $N_a$ has the property
of being a FINE polynomial, that is, in each Hepp sector the orderings of
the $\alpha $'s induce one and only one dominant monomial of $N$ \cite{hepp}%
. The point is that not all polynomials have the property of being FINE.
Nevertheless, there always exists the solution of splitting the $N$
polynomial in all its monomials, which will always be FINE. In other words,
if we adopt this extreme point of view, each piece $N_a$ is a monomial of $N$%
, which is necessarily FINE. In general, if the pieces $N_a$ are not the
monomials of $N$, they will generate a `multiple Mellin' representation of
the amplitude \cite{malbouisson7}. In the extreme situation of splitting $N$
in all its monomials, $N=\sum_KN_K$, a `complete Mellin' representation will
be generated \cite{malbouisson}. This is the point of view that we adopt
here.

To proceed, let us remember the following theorem \cite{courant}: for a
function $f(u)$, piecewise smooth for $u>0$, if the integral 
\begin{equation}
g(x)=\int_0^\infty du\,u^{-x-1}f(u)
\end{equation}
is absolutely convergent for $\alpha <$Re $x<\beta $, then 
\begin{equation}
f(u)=\frac 1{2\pi i}\int_{\sigma -i\infty }^{\sigma +i\infty }dx\,g(x)\,u^x,
\end{equation}
with $\alpha <\sigma <\beta $.

Then, if we take $u=N_K/U$ in the above theorem, it is easy to see that 
\begin{equation}
e^{-N_K/U}=\int_{\tau _K}\Gamma (-y_K)\left( \frac{N_K}U\right) ^{y_K},
\label{gamayk}
\end{equation}
where $\int_{\tau _K}$ is a short notation for $\int_{-\infty }^{+\infty }%
\frac{d(\text{Im }y_K)}{2\pi i}$, with Re $y_K$ fixed at $\tau _K<0$. We may
now recall the identity \cite{gradzhteyn} 
\begin{equation}
\Gamma (u)\left( A+B\right) ^{-u}=\int_{-\infty }^\infty \frac{d(\text{Im }x)%
}{2\pi i}\Gamma (-x)A^x\Gamma (x+u)B^{-x-u}  \label{gamauamaisb}
\end{equation}
and let us take $A\equiv U_1(x)$ and $B\equiv U_2+U_3+\cdots $. Using
iteratively the identity above, it can be shown that, for $u=\sum_Ky_K+d/2$, 
\begin{equation}
\Gamma \left( \sum_Ky_K+\frac d2\right) U^{-\sum_Ky_K-\frac d2}=\int_\sigma
\prod_j\Gamma (-x_j)U_j^{x_j},  \label{gamayk2}
\end{equation}
with Re $x_j=\sigma _j<0$, Re $\left( \sum_Ky_K+\frac d2\right) =\sum_K\tau
_K+\frac d2>0$, and $\int_\sigma $ means $\int_{-\infty }^{+\infty }\prod_j%
\frac{d(\text{Im }x_j)}{2\pi i}$ with $\sum_jx_j+\sum_Ky_K=-\frac d2$. Then,
after replacing eqs. (\ref{gamayk}) and (\ref{gamayk2}) in eq. (\ref
{amplitude2}), the amplitude is written as 
\begin{equation}
I_G(s_K,m_l^2)=\int_\Delta \frac{\prod_j\Gamma (-x_j)}{\Gamma (-\sum_jx_j)}%
\prod_Ks_K^{y_K}\Gamma (-y_K)\int_0^\infty \prod_l\,d\alpha _l\,\alpha
_l^{\phi _l-1}\,e^{-\sum_l\alpha _lm_l^2},  \label{equacao}
\end{equation}
where 
\begin{equation}
\phi _l\equiv \sum_ju_{lj}\sigma _j+\sum_Kn_{lK}\tau _K+1
\end{equation}
and $\Delta $ is the nonempty convex domain ($\sigma _j$ and $\tau _K$
standing respectively for $\text{Re }(-x_j)$ and $\text{Re }(-y_K)$), 
\begin{equation}
\Delta =\left\{ \sigma ,\tau \left| 
\begin{array}{l}
\sigma _j<0;\;\tau _K<0;\;\sum_jx_j+\sum_Ky_K=-\frac d2; \\ 
\forall i,\;\text{Re }\phi _l\equiv \sum_ju_{lj}\sigma _j+\sum_Kn_{lK}\tau
_K+1>0
\end{array}
\right. \right\}  \label{domain}
\end{equation}
and the symbol $\int_\Delta $ means integration over the independent
variables $\frac{\text{Im }x_j}{2\pi i}$, $\frac{\text{Im }y_K}{2\pi i}$.

The $\alpha $ integrations may be performed, using the well-known
representation for the gamma function, so that we have 
\begin{equation}
\int_0^\infty d\alpha _l\,e^{-\alpha _lm_l^2}\alpha _l^{\phi _l-1}=\Gamma
\left( \phi _l\right) \left( m_l^2\right) ^{-\phi _l}  \label{alphaint}
\end{equation}
and we finally get the complete Mellin representation of the amplitude in
the scalar case: 
\begin{equation}
I_G(s_K,m_l^2)=\int_\Delta \frac{\prod_j\Gamma (-x_j)}{\Gamma (-\sum_jx_j)}%
\prod_Ks_K^{y_K}\Gamma (-y_K)\prod_l\left( m_l^2\right) ^{-\phi _l}\Gamma
\left( \phi _l\right) .  \label{Mellin1}
\end{equation}
One should emphasize that the integral in eq.~ (\ref{equacao}) above is
assumed for simplicity to be ultraviolet convergent, but if this is not the
case, it is shown in \cite{malbouisson,rivasseau} that the renormalization
procedure does not alter the algebraic structure of integrands of the CM
representation. It only changes the set of relevant integration domains in
the Mellin variable space, all what follows remaining valid.

\section{The Complete Mellin representation and asymptotic behaviors for
gauge theories}

Turning now to gauge theories, we see that in both equations for $I_G(\xi
=1) $ and $I_G(\xi )$, given by eqs.~ (\ref{IG1novo}) and (\ref{IG2novo}),
the integrals over the Schwinger parameters have the same form as in the
scalar case in terms of the Symanzik polynomials, except for taking into
account the substitutions $m_i^2\rightarrow \bar{m}_i^2$ and $s_K\rightarrow 
\bar{s}_K$ and having to perform the derivatives on the auxiliary variables $%
\zeta $ which modify $m_i^2$ and $s_K$ and are put to zero at the end of the
calculation. Here $m_l$ stands generically for both the fermion masses and
the infrared gauge-field regulator. The derivatives are always attached to
Schwinger parameters, which then alter the power $\phi _l$ in the
integrations corresponding to (\ref{alphaint}), a different alteration for
each monomial of the $\mathcal{P}$ polynomial in the Feynman-gauge case, or
of the $\mathcal{P}$ and $\mathcal{Q}_1$ and $\mathcal{Q}_2$ polynomials for 
$I_G(\xi )$.

For $I_G(\xi =1)$ in eq.~ (\ref{IG1novo}), let us assume that the polynomial 
$\mathcal{P}$ has the form 
\begin{equation}
\mathcal{P}\left( -\frac 1{\alpha _i}\frac \partial {\partial \zeta
_i}\right) =\sum_Ac_A\left( -\frac 1{\alpha _i}\frac \partial {\partial
\zeta _i}\right) ^{d_i^A}.
\end{equation}
Then, according to \cite{rivasseau}, for each monomial $\prod_i\left( -\frac
1{\alpha _i}\frac \partial {\partial \zeta _i}\right) ^{d_i^A}$ of the
polynomial $\mathcal{P}$ the following complete Mellin representation holds: 
\begin{equation}
F_A^{(\xi =1)}(s_K,m_l^2)=\int_{\Delta _A}\left. \prod_i\left( -\frac
\partial {\partial \zeta _i}\right) ^{d_i^A}\frac{\prod_j\Gamma (-x_j)}{%
\Gamma (-\sum_jx_j)}\prod_K\bar{s}_K^{y_K}\Gamma (-y_K)\prod_l\left( \bar{m}%
_l^2\right) ^{-\phi _l+d_l^A}\Gamma \left( \phi _l-d_l^A\right) \right|
_{\zeta _l=0},  \label{Mellin0}
\end{equation}
where $\bar{m}_l$ stands for both the fermionic masses $\bar{m}_i$ and the
infrared regulator $\mu $ of the gauge-field propagators; for the latter
ones, $d_l^A=0$. Also, $\Delta _A$ is the modified nonempty convex domain 
\begin{equation}
\Delta _A=\left\{ \sigma ,\tau \left| 
\begin{array}{l}
\sigma _j<0;\;\tau _K<0;\;\sum_jx_j+\sum_Ky_K=-\frac d2; \\ 
\forall l,\;\text{Re }(\phi _l-d_l^A)\equiv \sum_ju_{lj}\sigma
_j+\sum_Kn_{lK}\tau _K-d_l^A+1>0
\end{array}
\right. \right\} .  \label{domain1}
\end{equation}
We see that now the derivatives may be taken out of the integration, so that
we define 
\begin{equation}
F_A^{(\xi =1)}(s_K,m_l^2)=\left. \prod_l\left( -\frac \partial {\partial
\zeta _l}\right) ^{d_l^A}F_A^{(\xi =1)}(\bar{s}_K,\bar{m}_l^2)\right|
_{\zeta _l=0},  \label{f1A}
\end{equation}
with 
\begin{equation}
F_A^{(\xi =1)}(\bar{s}_K,\bar{m}_l^2)=\int_{\Delta _A}\frac{\prod_j\Gamma
(-x_j)}{\Gamma (-\sum_jx_j)}\prod_K\bar{s}_K^{y_K}\Gamma (-y_K)\prod_l\left( 
\bar{m}_l^2\right) ^{-\phi _l+d_l^A}\Gamma \left( \phi _l-d_l^A\right) .
\label{Mellin2}
\end{equation}

In this case, a general asymptotic regime is defined by scaling the
invariants $\bar{s}_K$ and $\bar{m}_l^2$, 
\begin{eqnarray}
\bar{s}_K &\rightarrow &\lambda ^{a_K}\bar{s}_K,  \nonumber \\
\bar{m}_l^2 &\rightarrow &\lambda ^{a_l}\bar{m}_l^2,  \label{scaling}
\end{eqnarray}
where $a_K$ and $a_l$ may have positive, negative, or null values, and
letting $\lambda $ go to infinity. We then obtain under this scaling 
\begin{equation}
F_A^{(\xi =1)}(\lambda ,\bar{s}_K,\bar{m}_l^2)=\int_{\Delta _A}\left. \frac{%
\prod_j\Gamma (-x_j)}{\Gamma (-\sum_jx_j)}\prod_K\bar{s}_K^{y_K}\Gamma
(-y_K)\prod_l\left( \bar{m}_l^2\right) ^{-\phi _l+d_l^A}\Gamma \left( \phi
_l-d_l^A\right) \lambda ^{\psi ^A}\right| _{\zeta _l=0},  \label{Mellin3}
\end{equation}
where the exponent of $\lambda $ is the following linear function of the
Mellin variables: 
\begin{equation}
\psi ^A=\sum_Ka_Ky_K-\sum_la_l\left[ \phi _l(x_{j,}y_K)-d_l^A\right] .
\end{equation}

Then the proof of the theorem given in \cite{malbouisson7} for functions of
the form (\ref{Mellin1}) can be extended for (\ref{Mellin2}), and the
following theorem is valid: as $\lambda \rightarrow \infty $, we have an
asymptotic expansion of the integral $F_A^{(\xi =1)}(\lambda ,\bar{s}_K,\bar{%
m}_l^2)$ of the form 
\begin{equation}
F_A^{(\xi =1)}(\lambda ,\bar{s}_K,\bar{m}_l^2)=\sum_{p=p_{\text{max}%
}}^{-\infty }\sum_{q=0}^{q_{\text{max}}(p)}F_{pq}^{(\xi =1)A}(\bar{s}_K,\bar{%
m}_l^2)\lambda ^p\ln ^q\lambda ,  \label{Mellin4}
\end{equation}
where $p$ runs over the rational values of a decreasing arithmetic
progression, with $p_{\text{max}}$ as a `leading power', and $q$, for a
given $p$, runs over a finite number of nonnegative integer values.

A systematic way of evaluating the coefficients $F_{pq}$ can be done by
successive analytical continuations of the linear form $\psi $ to
``smaller'' cells of the type $\Delta $ in eq.~ (\ref{domain}). Under the
scaling (\ref{scaling}), and summing over all monomials, $I_G(\xi =1)$ in
eq. (\ref{IG1novo}) becomes 
\begin{eqnarray}
I_G\left( \xi =1,\lambda \right)  &=&C_G\sum_Ac_A\prod_l\left( -\frac
\partial {\partial \zeta _l}\right) ^{d_l^A}\left. \sum_{p=p_{\text{max}%
}}^{-\infty }\sum_{q=0}^{q_{\text{max}}(p)}\left[ F_{pq}^{(\xi =1)A}(\bar{s}%
_K,\bar{m}_l^2)\right] \lambda ^p\ln ^q\lambda \right| _{\zeta _l=0} 
\nonumber \\
&=&C_G\left. \sum_{p=p_{\text{max}}}^{-\infty }\sum_{q=0}^{q_{\text{max}%
}(p)}\left[ \sum_Ac_A\prod_l\left( -\frac \partial {\partial \zeta
_l}\right) ^{d_l^A}F_{pq}^{(\xi =1)A}(\bar{s}_K,\bar{m}_l^2)\right] \lambda
^p\ln ^q\lambda \right| _{\zeta _l=0}  \nonumber \\
&=&C_G\sum_{p=p_{\text{max}}}^{-\infty }\sum_{q=0}^{q_{\text{max}%
}(p)}G_{pq}^{(\xi =1)}(s_K;m_l^2)\lambda ^p\ln ^q\lambda ,  \label{expansao1}
\end{eqnarray}
where we have defined 
\begin{equation}
G_{pq}^{(\xi =1)}(s_K;m_l^2)=\sum_Ac_A\left. \prod_l\left( -\frac \partial
{\partial \zeta _l}\right) ^{d_l^A}F_{pq}^{(\xi =1)A}(\bar{s}_K,\bar{m}%
_l^2)\right| _{\zeta _l=0}.  \label{Gpq1}
\end{equation}

For $I_G(\xi )$, generalizing what has been done for eq.~ (\ref{Mellin0}),
we consider the product of polynomials $\mathcal{P}$, $\mathcal{Q}_1$ and $%
\mathcal{Q}_2$ in eq. ~(\ref{IG2novo}) in the form 
\begin{equation}
\mathcal{PQ}_1\mathcal{Q}_2=\sum_Bc_B(\xi )\prod_i\left( -\frac 1{\alpha
_i}\frac \partial {\partial \zeta _i}\right) ^{d_i^B}\left( -\frac 1{\beta
_{r_1}}\frac \partial {\partial \zeta _{r_1}}\right) ^{e_{r_1}^B}\left(
-\frac 1{\beta _{r_2}}\frac \partial {\partial \zeta _{r2}}\right)
^{e_{r_2}^B}.  \label{pq1q2}
\end{equation}
In order to proceed, notice that eq. ~(\ref{sym}) is modified in the present
case to 
\begin{eqnarray}
U(\alpha ,\beta ) &=&\sum_j\prod_{i\in F}\alpha _i^{u_{ij}}\prod_{r_1\in
\Gamma }\beta _{r_1}^{v_{r_1j}}\prod_{r_2\in \Gamma }\beta _{r_2}^{v_{r_2j}}
\label{ualphabeta} \\
N(\alpha ,\beta ) &=&\sum_K\bar{s}_K\prod_{i\in F}\alpha
_i^{n_{ij}}\prod_{r_1\in \Gamma }\beta _{r_1}^{m_{r_1K}}\prod_{r_2\in \Gamma
}\beta _{r_2}^{m_{r_2K}}.
\end{eqnarray}
We now replace the above expressions in eq.~ (\ref{IG2novo}), together with
eqs.~ (\ref{gamayk}) and (\ref{gamayk2}). After some manipulations, we find 
\begin{eqnarray}
I_G(\xi ) &=&C_G\sum_Bc_B(\xi )\int_{\Delta _B}\frac{\prod_j\Gamma (-x_j)}{%
\Gamma \left( -\sum_jx_j\right) }\prod_K\Gamma (-y_K)  \nonumber \\
&&\times \left\{ \prod_i\left( -\frac \partial {\partial \zeta _i}\right)
^{d_i^B}\int d\alpha _i\,e^{-\sum_i\alpha _i\bar{m}_i^2}\,\alpha _i^{\phi
_i-d_i^B-1}\right.   \nonumber \\
&&\times \prod_{r_1}\left( -\frac \partial {\partial \zeta _{r_1}}\right)
^{e_{r_1}^B}\int d\beta _{r_1}\,e^{-\sum_{r_1}\beta _{r_1}\bar{\mu}%
_{r_1}^2}\,\alpha _i^{\phi _{r_1}-e_{r_1}^B-1}  \nonumber \\
&&\times \left. \left. \prod_{r_2}\left( -\frac \partial {\partial \zeta
_{r_2}}\right) ^{e_{r_2}^B}\int d\beta _{r_2}\,e^{-\sum_{r_2}\beta _{r_2}%
\bar{\mu}_{r_2}^2}\,\alpha _i^{\phi _{r_2}-e_{r_2}^B-1}\right\} \bar{s}%
_K^{y_K}\right| _{\zeta =0},
\end{eqnarray}
where $\phi _i=\sum_ju_{ij}x_j+\sum_Kn_{iK}y_K$ as before, and 
\begin{equation}
\phi _{r_{1,2}}=\sum_jv_{jr_{1,2}}x_j+\sum_Km_{Kr_{1,2}}y_K
\end{equation}
and now the domain of integration is 
\begin{equation}
\Delta _B=\left\{ \sigma ,\tau \left| 
\begin{array}{l}
\sigma _j<0;\;\tau _K<0;\;\sum_jx_j+\sum_Ky_K=-\frac d2; \\ 
\forall i,\text{Re }(\phi _i-d_i^B)\equiv \sum_ju_{ij}\sigma
_j+\sum_Kn_{iK}\tau _K-d_i^B+1>0; \\ 
\forall r_1,r_2,\text{Re }(\phi _{r_{1,2}}-e_{r_{1,2}}^B)\equiv
\sum_jv_{jr_{1,2}}\sigma _j+\sum_Km_{Kr_{1,2}}\tau _K-e_{r_{1,2}}^B+1>0
\end{array}
\right. \right\} .
\end{equation}
The integrals over all Schwinger parameters $\alpha $ and $\beta $ are
performed in the same fashion as in eq.~ (\ref{alphaint}), so the amplitude
is written as 
\begin{eqnarray}
I_G(\xi ) &=&C_G\sum_Bc_B(\xi )\prod_i\left( -\frac \partial {\partial \zeta
_i}\right) ^{d_i^B}\prod_{r_1}\left( -\frac \partial {\partial \zeta
_{r_1}}\right) ^{e_{r_1}^B}\prod_{r_2}\left( -\frac \partial {\partial \zeta
_{r_2}}\right) ^{e_{r_2}^B}  \nonumber \\
&&\times \int_{\Delta _B}\frac{\prod_j\Gamma (-x_j)}{\Gamma (-\sum_jx_j)}%
\prod_K\bar{s}_K^{y_K}\Gamma (-y_K)  \nonumber \\
&&\times \left. \Gamma (\phi _i-d_i^B)\left( \bar{m}_i^2\right) ^{-\phi
_i+d_i^B}\Gamma (\phi _{r_1}-e_{r_1}^B)\left( \bar{\mu}_{r_1}^2\right)
^{-\phi _{r_1}+e_{r_1}^B}\Gamma (\phi _{r_2}-e_{r_2}^B)\left( \bar{\mu}%
_{r_2}^2\right) ^{-\phi _{r_2}+e_{r_2}^B}\right| _{\zeta =0},  \label{IG2IG2}
\end{eqnarray}
where $j$ runs over the whole set of monomials of $U(\alpha ,\beta )$, eq. (%
\ref{ualphabeta}). Then, performing steps similar to those that led to eq. (%
\ref{f1A}) for $I_G(\xi =1)$, we get for each monomial of the product of
polynomials $\mathcal{PQ}_1\mathcal{Q}_2$ in eq. (\ref{pq1q2}) the
expression 
\begin{equation}
F_B(s_K,m_i^2,\mu ^2)=\left. \prod_i\left( -\frac \partial {\partial \zeta
_i}\right) ^{d_i^B}\prod_{r_1}\left( -\frac \partial {\partial \zeta
_{r_1}}\right) ^{e_{r_1}^B}\prod_i\left( -\frac \partial {\partial \zeta
_{r_2}}\right) ^{e_{r_2}^B}F_B(\bar{s}_K,\bar{m}_i^2,\bar{\mu}%
_{r_{1,2}}^2)\right| _{\zeta =0},
\end{equation}
where 
\begin{equation}
F_B(\bar{s}_K,\bar{m}_i^2,\bar{\mu}_{r_{1,2}}^2)=\int_{\Delta _B}\frac{%
\prod_j\Gamma (-x_j)}{\Gamma (-\sum_jx_j)}\prod_K\bar{s}_K^{y_K}\Gamma
(-y_K)\prod_l\left( \bar{m}_l^2\right) ^{-\phi _l+d_l^B}\Gamma \left( \phi
_l-d_l^B\right) .  \label{f2B}
\end{equation}

We now consider an asymptotic regime defined by 
\begin{eqnarray}
\bar{s}_K &\rightarrow &\lambda ^{a_K}\bar{s}_K,  \nonumber \\
\bar{m}_i^2 &\rightarrow &\lambda ^{a_i}\bar{m}_i^2,  \nonumber \\
\bar{\mu}_{r_{1,2}}^2 &\rightarrow &\lambda ^{a_{r_{1,2}}}\bar{\mu}%
_{r_{1,2}}^2  \label{scaling1}
\end{eqnarray}
where $a_K$, $a_i$ and $a_{r_{1,2}}$ may have positive, negative, or null
values, and we let $\lambda $ go to infinity. Under the above scaling, $F_B$
in eq. (\ref{f2B}) then becomes a function of $\lambda $, given by 
\begin{equation}
F_B(\bar{s}_K,\bar{m}_i^2,\bar{\mu}_{r_{1,2}}^2)=\int_{\Delta _B}\frac{%
\prod_j\Gamma (-x_j)}{\Gamma (-\sum_jx_j)}\prod_K\bar{s}_K^{y_K}\Gamma
(-y_K)\prod_l\left( \bar{m}_l^2\right) ^{-\phi _l+d_l^B}\Gamma \left( \phi
_l-d_l^B\right) \lambda ^{\psi ^B},  \label{f2Bmod}
\end{equation}
with the exponent of $\lambda $ being 
\begin{equation}
\psi
^B=\sum_Ka_Ky_K-\sum_ia_i(x_j,y_K)-d_i^B-%
\sum_{r_1}a_{r_1}(x_j,y_K)-e_{r_1}^B-\sum_{r_2}a_{r_2}(x_j,y_K)-e_{r_2}^B.
\end{equation}
Since the integral in the function (\ref{f2Bmod}) has the same form of that
of the corresponding one in the scalar field theory, the following
asymptotic behavior expansion holds: 
\begin{equation}
F_B(\lambda ,\bar{s}_K,\bar{m}_i^2,\bar{\mu}_{r_{1,2}}^2)=\sum_{p=p_{\text{%
max}}}^{-\infty }\sum_{q=0}^{q_{\text{max}}(p)}F_{pq}^B(\bar{s}_K,\bar{m}%
_i^2,\bar{\mu}_{r_{1,2}}^2)\lambda ^p\ln ^q\lambda .  \label{f2Bexp}
\end{equation}
Therefore, from (\ref{f2Bmod}), 
\begin{equation}
I_G(\lambda ,\xi )=C_G\sum_{p=p_{\text{max}}}^{-\infty }\sum_{q=0}^{q_{\text{%
max}}(p)}G_{pq}(\xi ;s_K^{(2)};m_i^2,\mu ^2)\lambda ^p\ln ^q\lambda ,
\end{equation}
where 
\begin{equation}
G_{pq}(\xi ;s_K;m_i^2,\mu ^2)=\sum_Bc_B(\xi )\left. \prod_i\left( -\frac
\partial {\partial \zeta _i}\right) ^{d_i^B}\prod_{r_1}\left( -\frac
\partial {\partial \zeta _{r_1}}\right) ^{e_{r_1}^B}\prod_i\left( -\frac
\partial {\partial \zeta _{r_2}}\right) ^{e_{r_2}^B}F_{pq}^B(\bar{s}_K,\bar{m%
}_i^2,\bar{\mu}_{r_{1,2}}^2)\right| _{\zeta =0}.  \label{Gpq2}
\end{equation}

\section{Concluding remarks}

We have  stated a theorem which generalizes for gauge field theories in an
unspecified gauge previous results on the asymptotic behaviors of scalar
field theories. Its relevance stems from physical situations
in which the choice of gauges other than the Feynman gauge is mandatory.
These situations do exist, as we have mentioned in the introduction.
The proof of the theorem has been possible by conveniently changing the
invariants $s_K$, $m_i^2$ and $\mu ^2$ into new objects $\bar{s}_K$, $\bar{m}%
_i^2$ and $\bar{\mu}_{r_{1,2}}^2$ by means of dummy variables $\zeta $
associated to each internal line of the diagram, to be taken equal to zero
in the end. It results that the relevant Feynman integrals are expressed as
polynomials of derivatives with respect to this dummy variables, acting on
integrals over the Schwinger parameters. Complete Mellin representations can
be written for gauge theory amplitudes and a previous theorem for asymptotic
behaviors of scalar amplitudes was generalized for them. It results that the
dependence on the gauge-fixing parameter is contained in the coefficients of
the asymptotic expansion.

We recall that the idea of the complete Mellin representation is an extreme version
of the multiple Mellin' representation \cite{malbouisson7}. 
In this case, the polynomial $N$  is split into FINE
pieces,  that is, in each Hepp sector the orderings of
the Schwinger variables induce one and only one dominant monomial of $N$. 
The fact is that not all polynomials have the property of being FINE, and in some cases,
the only solution is to split the $N$
polynomial in all its monomials, which will always be FINE, and generates the
complete Mellin representation. 

A point to be emphasized 
is the following. One may wonder why such an
involved method should be used in order to obtain an asymptotic expansion of
Feynman amplitudes, at the price of employing a rather cumbersome notation,
instead of using the much simpler technique of the Mellin transform. The
answer is that the Mellin transform method, which is equivalent to a simple
Mellin representation, is valid only if the scaled part of the Symanzik
polynomial $N$ has the property of being FINE. This is
the case for simple (low-order) Feynman diagrams, but it is not true in
general for an arbitrary diagram. For all those high-order diagrams, whose inner complexity leads
to a non-FINE $N$ polynomial, and for which the Mellin transform (or, equivalently, the
simple Mellin representation) method does not apply,
our theorem guarantees the existence of the asymptotic 
expansions with the same structure as that of the simpler FINE cases.

\begin{center}
{\bf Acknowledgments}
\end{center}

This work received partial financial support from CNPq/MCT (Brazil). I.R. acknowledges partial financial support from Pronex/FAPERJ.

\end{document}